\documentclass[12pt,aps,prb,preprint,showpacs]{revtex4}
\usepackage{bm}
\input{epsf}
\begin{document}
\title{Optical properties of nanocrystalline Y$_{2}$O$_{3}$:Eu$^{3+}$}
\author{S. Ray}
\author{P. Pramanik} 
\affiliation{Department of Chemistry, Indian Institute of 
Technology,\\ Kharagpur 721 302, West Bengal, India}
\author{A. Singha}
\author{Anushree Roy$^{\mbox a)}$}
\affiliation{Department of Physics and Meteorology, 
Indian Institute of Technology,\\ Kharagpur 721 302, West Bengal, India}
\begin{abstract}

Optical properties of nanocrystalline red-emitting phosphor, Europium
doped Yttria (Y$_{2}$O$_{3}$:Eu$^{3+})$, of average particle size 15 nm
are investigated. The intensity of the strongest emission line 
at 612
nm is found to be highest in the nanocrystalline sample with 4 at.  wt.
$\%$ of Europium. The narrow electronic emission spectrum
suggests a crystalline surrounding in this nanomaterial. We have estimated
the strength of the crystal field parameter at the dopant
site, which plays a crucial role in determining the appearance of the
intense emission line. The equilibrium temperature of this
system has also been calculated from the intensity ratio of Stokes and
anti-Stokes Raman scattering. Though known for the bulk samples, our
approach and consequent results on the crystalline nanomaterial of
Y$_{2}$O$_{3}$:Eu$^{3+}$ provide a unique report, which, we believe, can
be of considerable significance in nanotechnology.  
\end{abstract}
\maketitle
\def\d{{\mathrm{d}}}
\section{Introduction}

The optical properties of rare-earth ions trapped in inorganic oxides
continue to be a research attraction in terms of both their fundamental
and technological importance.\cite{1,2,3,4,5} Yttrium oxide doped with
Eu$^{3+}$ is one of the main red-emitting phosphors and is
widely used in lighting industry and in solid-state-laser based
devices.\cite{1,2,3} The unique optical properties of this material is
based on the $f$- and $d$ electrons of Europium ions.  To enhance the
brightness and resolution of displays in devices in the present
nanotechnology regime, it is important to develop phosphors with
controlled morphology and small particle size (nanoparticles). The
potential of this low dimensional material in fabrication of modern
micro/nano devices requires an understanding of their fundamental
properties in detail. In particular, the origin and behavior of the
strongest ${}^{5}D_{0}-{}^{7}F_{2}$ transition spectrum from Eu$^{3+}$
needs to be investigated because of its importance in designing laser
devices. For example, permanent laser-induced gratings
have been fabricated using crossed write beams in resonance
with the ${}^{7}F_{2}-{}^{5}D_{0}$ absorption transition of
Eu$^{3+}$.\cite{6}

Over the last couple of decades, a considerable amount of work
on the spectral properties of bulk
Y$_{2}$O$_{3}$:Eu$^{3+}$ has been reported in the
literature.\cite{7,8,9,10,11,12,13,14,15}. Additionally, we find several
articles on nanoparticles of this material, though, the
detailed study of their optical properties is, as far as
we are aware, incomplete. In recent times, Schmechel {\em et
al.}\cite{16} have reported luminescence properties of nanocrystalline
Y$_{2}$O$_{3}$:Eu$^{3+}$. However, due to the presence of
defect/disordered states in this low dimensional system, the quantum
efficiency of this nanomaterial is much less than the corresponding bulk
commercial sample. Furthermore, Konard {\em et al.}\cite{17}
have synthesized and studied the luminescence properties of
Y$_{2}$O$_{3}$:Eu$^{3+}$ nanoparticles, though they do not discuss the
quantum efficiency of this material. In both these studies \cite{16,17}
one does not find the details of the optical emission spectrum of this low
dimensional system.

Apart from optical properties, other characteristics of such
inorganic, lanthanide-doped oxides also play an important role in using
these materials in technology. Walsh {\em et al.}{\cite{18} have recently
demonstrated that the strength of the crystal field is crucial in
improving the performance of solid-state lanthanide lasers. It may be
noted that the doped Europium ions in these oxides provide the high-energy
local mode phonons, which can produce structural modification of the host. 
Finally, the knowledge on the thermal properties of this
material is required for efficient laser operation (for example, thermal
lensing) as well as the stability of laser-induced gratings. Such
gratings, designed from Europium ion doped glasses, are found to be stable
up to a certain equilibrium temperature, but get erased with rising
thermal energy.\cite{6}
 
In this article, we have addressed optical properties of nanocrystalline
Europium doped Yttria, prepared by soft chemical route. Section II covers
the sample preparation and characterization of the nanomaterial by x-ray
diffractometry and Transmission Electron microscopy (TEM). In Section III,
we have reported the electronic emission spectrum for the range of 490 nm
- 900 nm by Photoluminescence spectroscopy. The crystal field effect
and equilibrium temperature in the nanocrystalline
Y$_{2}$O$_{3}$:Eu$^{3+}$ have been discussed in Section IV. Finally, in
Section V we have summarized our results with a few concluding remarks.

\section{SAMPLE PREPARATION AND CHARCTERIZATION}

The nanomaterial has been prepared by soft-chemical method using TEA
(Triethanolamine) as complexing (chelating) agent.\cite{19} Solid powder
of Europium Oxide and Yttrium Oxide (both from ALDRICH) are boiled with
concentrated nitric acid in water bath to get clear solutions of Europium
Nitrate and Yttrium Nitrate, respectively. The addition of nitric acid is
continued until the solution reached a pH of 3.0. Europium 
Nitrate and
Yttrium Nitrate solutions are taken in different stoichiometric 
ratios in
order to vary the Eu$^{3+}$ concentration by 1-7 at. wt. 
$\%$ with respect
to Y$^{3+}$. These nitrate solutions and TEA are taken in a beaker and
then heated on a hot plate at a temperature of about 180-200 
$^\circ$C.  
TEA
helps to keep the metal ions in homogenous solution through the reaction
without precipitation. Due to complete dehydration of the complex
precursor and decomposition of the metal complexes upon heating the
mixture, we get a fluffy mass. It is calcined at 900 $^\circ$C for four
hours.  A white powder is obtained. In  the rest of this 
article,  we denote the nanomaterial of Y$_2$O$_3$ doped with 4 at. wt. 
$\%$ Eu$^{3+}$ (in
stoichiometric ratio) as Sample A. The micron-sized particles (Sample B)
are prepared by heating a part of the above sample at 1200 $^\circ$C for
six hours, after which white bulk Y$_{2}$O$_{3}$:Eu$^{3+}$ phosphor is
obtained.

Samples for Transmission Electron Microscopy are deposited onto 300
mesh copper TEM grids coated with 50 nm carbon films. Nanocrystals,
dissolved in Acetone, are placed on the grid drop-wise. The excess liquid
is allowed to evaporate in air. The grids are examined in 

\newpage

\noindent
JEOL 2010
microscope with Ultra-High Resolution (UHR) pole-piece using a LaB$_6$
filament operated at 200 kV.  The high resolution transmission electron
microscopy (HRTEM) image for Sample A, shown in Fig. 1(a), clearly
demonstrates the lattice fringes for the cubic phase of Y$_{2}$O$_{3}$:
Eu$^{3+}$. From the micrograph we have determined the lattice spacing in
the particle to be 0.29 nm, which corresponds to (222) plane of cubic
phase of this material. The frequency plot of the size distribution, shown
in Fig. 1(b), is obtained by measuring the size of many 
particles per
sample. Size distribution for the nanoparticles are usually found to be
log-normal:\cite{20}

\begin{equation}
P(d)=\frac{1}{d\sigma\sqrt{2\pi}}exp\left(-\frac{ln^{2} 
(d/\bar{d})}{2\sigma^{2}}\right).
\end{equation}                                  

\noindent 
Here $\bar{d}$ and $\sigma$ are related to the average size and
the size distribution of the particles. By fitting the frequency plot
using Eq. (1) [solid line in Fig. 1(b)], the average particle size of
Sample A has been estimated to be 15 nm with a very narrow size
distribution ($\sigma$ = 0.14).  The average particle size of Sample B is
found to be 500 nm with a wider size distribution.


Samples have been characterized by x-ray diffraction at room temperature 
using Philips PW1710 X-ray Diffractometer, equipped with a vertical 
goniometer and CuK$_\alpha$ radiation source of wavelength 
($\lambda$=1.514 \AA). X-ray 
diffraction patterns for both Sample A and B are shown in Fig. 2. The 
observed values of the diffraction angles (2$\theta$) match quite well 
with the 
reported data and the patterns exhibit cubic symmetry (space group 
$Ia$3).\cite{8,21} Identical diffraction patterns of both Sample A and 
Sample B 
indicate that the lattice constants in both the materials are same 
{\em i.e.} 
there is no distortion in the structure of Sample A because of lowering of 
dimension of the system. We have calculated the average nanocrystallite 
size ($L$)  using Scherrer's equation\cite{22}

\begin{equation}
\Gamma=\frac{0.94\lambda}{L\cos\theta},
\end{equation}					

\noindent
where $\Gamma$ (expressed in radians) is the full width at the half maxima 
(FWHM) of the x-ray diffraction peak. $\theta$ is the diffraction peak 
position. Taking 
the diffraction resolution function (DRF) to be Gaussian and the 
diffraction line to be Lorentzian, the line-width correction for the 
x-ray 
instrumental broadening can be taken into account by using the empirical 
expression \cite{23}

\begin{equation}
\eta=\eta_{e}-\left(\frac {1}{\eta_{e}}\right)^{\beta},
\end{equation}


\noindent				      
where, $\beta=1-0.1107/\eta_{e}$. $\eta_{e}$ is the ratio of observed 
FWHM ($\Gamma_{e}$) and the FWHM ($\Gamma_{d}$) of the 
DRF, whereas, $\eta$ is the ratio of the true FWHM ($\Gamma$) of the 
diffraction line 
and $\Gamma_{d}$. For our instrument $\Gamma_{d}$ = 0.10$^\circ$. For the 
diffraction peak at 2$\theta$ = 28.974$^\circ$ of the Sample A, 
$\Gamma_{e}$ = 0.326$^\circ$, which gives 
$\Gamma$ = 0.254$^\circ$. From Eq. (2), 
the crystallite size has been estimated, which is close to what we have 
observed from the HRTEM images of the particles, discussed earlier. The 
narrow diffraction pattern from the nanoparticles implies the crystalline 
nature of Sample A.

\section{OPTICAL EMISSION  SPECTRA}


Luminescence spectra are obtained using a 1200 g/mm 
holographic grating, a 
holographic supernotch filter, and a Peltier cooled CCD detector. 
{\bf We use a 488 nm Argon ion laser as our 
excitation source.}

The electronic transition spectra for the range 490 nm to 900 nm are shown
in Fig. 3. Many of these transitions satisfy the magnetic dipole selection
rules ($\Delta J$ = 0, $\pm$ 1 with $J$ = 0 $\not\leftrightarrow$ $J'$ = 
0) in an intermediate coupling scheme (shown in Fig. 4).\cite{24} 
{\bf In free
Eu$^{3+}$, the ${}^{5}D_{0}-{}^{7}F_{2}$ transition is 
forbidden (for both
magnetic and electric dipoles).} This holds true in a crystalline
environment, which has inversion symmetry. {\bf It is well 
established that in
the body centred cubic structure of Yttria, Eu$^{3+}$ replaces Y$^{3+}$ in 
two
different sites. 75$\%$ of these sites are with C$_{2}$ and the other
25$\%$ are with S$_6$ symmetry.} Between these two the former does not 
have
the inversion symmetry.\cite{25,26,27,28,29} According to the theory of
Judd\cite{30} and Ofelt \cite{31} the ${}^{5}D_{0}-{}^{7}F_{2}$
transition becomes electric dipole type, due to an admixture of opposite
parity $4f^{n-1}5d$ states by an odd parity crystal field component.  
{\bf ${}^{5}D_{0}-{}^{7}F_{3,5}$ transitions have a mixed character - 
magnetic
dipole intensity usually to first order and electric dipole intensity to
second order.\cite{32}} The ${}^{5}D_{0}-{}^{7}F_{3}$ transition exhibits
a relatively larger intensity {\bf than what is expected because of its 
large
induced electric dipole character.} The large oscillator strength of the
${}^{5}D_{0}-{}^{7}F_{3}$ transition may be attributed to significant
crystal-field induced mixing of the ${}^{7}F_{3}$ state to $^{7}F_{2}$
state, leading to the transfer of energy from ${}^{5}D_{0}-{}^{7}F_{2}$ to
${}^{5}D_{0}-{}^{7}F_{3}$ transition.\cite{33}  
${}^{5}D_{0}-{}^{7}F_{4,6}$ transitions are electric dipole in character
and have appreciable intensities.\cite{32} The magnetic dipole and
electric quadrupole transitions {\bf are forbidden for 
$J_{0}-J_{0}$
line.} The theory of Judd and Ofelt does not explain the
${}^{5}D_{0}-{}^{7}F_{0}$ fluorescence of Eu$^{3+}$ ion because the
$J_{0}-J_{0}$ transition is forbidden even in the presence of
non-centrosymmetric potential. The origin of this emission line may 
be} explained by the variation in crystal field potential, which causes 
the
mixing of odd parity $J$ = 1 states into the ${}^{7}F_{0}$ and 
${}^{5}D_{0}$ states.\cite{34} {\bf However,} from the fluorescence 
line-narrowing experiment, it has
been shown that the ${}^{5}D_{0}-{}^{7}F_{0}$ borrows its
intensity from ${}^{5}D_{0}-{}^{7}F_{2}$ line through the $J$-mixing in
this material.\cite{35} Other than the above transitions, direct band to 
band transition and excitonic emission appear at 6.2 eV\cite{36} and 
5.9 eV\cite{36}, respectively, in this system. {\bf We could not
study these emission lines because of our experimental limitations 
(recall that we have used 488 nm as the wavelength of our excitation 
source).} No bands are reported below 490 nm. However,
electronic transitions from ${}^{5}D_{3}$ to ${}^{7}F_{J}$ is known to 
occur in this region.\cite{7} In Fig. 4 we have shown the energy level 
diagram of the low-lying states of the Eu$^{3+}$ ion in nanocrystalline
Y$_{2}$O$_{3}$:Eu$^{3+}$, as obtained in our experiment.


It is well known that a strong energy transfer takes place from S$_6$ to 
C$_2$ 
site, which gets stronger with increase in Europium content.\cite{37} 
Consequently, the ratio of {\bf the emission} intensity of C$_2$ site to 
the 
emission 
intensity of S$_6$ site also increases with increase in Europium 
concentration in the sample.\cite{34} Previous emission study on cubic 
bulk Y$_{2}$O$_{3}$:Eu$^{3+}$  has shown that only  the transitions 
${}^{5}D_{0}-{}^{7}F_{1}$ originate from 
S$_6$ site,\cite{38,39} whereas nearly all of the other features in the 
electronic 
spectrum are due to  Eu$^{3+}$ ion in C$_2$ sites.   The different 
behaviour of Eu$^{+3}$ ion in two different lattice sites is most 
probably due to {\bf the large} 
diameter (0.95 \AA) of Eu$^{3+}$ ion, which fills the lattice sites of the 
host, and 
gets affected by nearby atoms.  Fig. 5 shows the change in ratio of the 
peak intensities of emission spectra for ${}^{5}D_{0}-{}^{7}F_{0}$ (581 
nm) transition of 
C$_{2}$ site and ${}^{5}D_{0}-{}^{7}F_{1}$ transitions (587 nm) of 
S$_6$ site for nanocrystalline 
Y$_{2}$O$_{3}$:Eu$^{3+}$ with different Europium ion 
concentration. Due to the 
energy transfer from S$_6$ site to C$_2$ site, the ratio increases till 
the 
dopant concentration is 4 at. wt. $\%$ of {\bf the host} material, but 
decreases 
with 
further increase in mole concentration of Europium ion. This drop in the 
ratio with concentration of Eu$^{3+}$ beyond 4 at. wt. $\%$ can be 
attributed 
to fluorescence quenching, which arises due to the energy transfer or 
electron transfer process between the two nearest excited Eu$^{3+}$ and 
dominates over the spontaneous emission process.\cite{40} 
In the above analysis we have 
compared ${}^{5}D_{0}-{}^{7}F_{1}$ transition of S$_6$ site with 
${}^{5}D_{0}-{}^{7}F_{0}$ transition of C$_2$ 
site, as these two lines are of comparable {\bf intensity. Justifying the 
above discussion, we now show} that the effect of fluorescence 
quenching becomes dominant in the case of the most intense emission line 
at 612 nm for the sample 
doped with more than 4 at. wt. $\%$ of Eu$^{3+}$. We have fitted the 
emission spectrum at 
612 nm for all the samples with Lorentzian line shape.  In the inset of 
Fig. 5 we have shown the variation in intensity {\bf of this emission 
spectrum} with concentration 
of Eu$^{3+}$ in nanocrystalline Y$_{2}$O$_{3}$. The sample with 
4 at. wt. $\%$ of 
dopant 
concentration (Sample A) shows highest emission intensity for 612 nm 
transition line. Hence, we have taken this particular material as a 
standard nanocrystalline sample and compared its physical properties with 
bulk (Y$_{2}$O$_{3}$:Eu$^{3+}$).


In Fig. 6, we have compared the most intense electronic transition line at
612 nm in the spectra for both nanomaterial (dashed-dotted line : Sample
A) and the bulk system (solid line : Sample B).  Both the spectra have
been fitted with three Lorentzian line shapes. In the non-linear least
square fit the width of the spectral line and its intensity have been kept
as varying parameters. We have not observed any shift in the luminescence
peak position due to quantum size effect. As shown in Fig. 6, the
luminescence intensity of 612 nm transition from Sample A is slightly less
than that observed in Sample B. Surface states are the sources of
non-radiative recombination centers. Due to high surface to volume ratio
in nanoparticles compared to the corresponding bulk material the
non-radiative recombination processes dominate in the former. Moreover,
there is a possibility of absorption by impurities in nanostructured
materials. These, in turn, decrease the luminescence efficiency in
nanoparticles.\cite{16} However, here we would like to point out that the
drop in intensity in nanocrystalline sample compared to the corresponding
bulk material, prepared by us, is much less than what has been reported
before in the literature,\cite{16} due to the absence of impurities and
defects in Sample A (as we have observed from our x-ray measurements).  
We note that (i) the half-widths of the 612 nm line for sample A and 
sample
B have been obtained as 0.5 $\pm$ 0.2 nm and 0.4 $\pm$ 0.2 nm,
respectively. The sharp transition spectrum for Sample A suggests nearly
crystalline surrounding and (ii) the appearance of all lines from 
${}^{5}D_{2}$ and ${}^{5}D_{1}$ signify absence of too
many non-radiative cross-relaxation processes in Sample A.


\section{CRYSTAL FIELD EFFECT AND EQUILIBRIUM TEMPERATURE}


The most intense forced electric dipole ${}^{5}D_{0}-{}^{7}F_{2}$
transition line, discussed in the previous section, is expected to be
hypersensitive to the crystal field at the dopant site. The crystal field
Hamiltonian can be written as\cite{41}

\begin{equation}
H_{CF}=\sum_{l,m}\sum_{i}B_{lm}C_{lm}(\theta_{i},\phi_{i}),
\end{equation}  			 				     

\noindent
where $B_{lm}$, the crystal field parameter, which describes the effect 
of the crystal field on the free ion Hamiltonian, satisfies 

\begin{equation}
B_{lm}=(-1)^{m}B_{l,-m}
\end{equation}

\noindent
and $C_{lm}$ are spherical tensors defined in terms of spherical 
harmonics according to

\begin{equation}
C_{lm}(i)=\left(\frac{4\pi}{2l+1}\right)^{1/2}Y_{l,m}(\theta_{i},\phi_{i}).
\end{equation} 

\noindent
In Eq. (4) - (6) $l$ runs over 2, 4, 6...... and $m$ = 0, $\pm$ 2, $\pm$ 
4...... $\pm l$. The sum on $i$ runs over the number of electrons in the 
4$f^n$ configuration. $\theta_i$ and $\phi_i$ 
are the angular coordinates of the $i$th electron. The reason for the 
presence of only even $l$ in the sum is related to the fact that even $l$ 
terms in the expansion of Eq. (4) contribute to the shifting and splitting 
of energy levels. On the other hand, the odd $l$ terms in Eq. (4), if 
included, would represent the mixing of opposite parity states from higher 
lying configuration into the 4$f^n$ one.  The usual analysis with 
the crystal field 
potential, in Eq. (4), does not take into account the mixing of different 
$J$-states. $l$=0 is always ignored because it corresponds to spherically 
symmetric crystal field that shifts all energy levels equally without affecting the 
energy level splitting. However, as discussed before, $J$-mixing 
plays an important role in explaining the optical emission spectrum of the 
material. Assuming that the $D_{0}-F_{0}$ transition is allowed only by 
$J$-mixing, the intensity variation of  ${}^5D_{0}-{}^{7}F_{0}$ line is 
ascribed to the effect of mixing of  ${}^{7}F_{2}$ into ${}^{7}F_{0}$ 
states. This can be taken into account 
by considering the second-order term (crystal field parameter $B_{20}$) in 
the above local field potential.\cite{33} The transition strength of  
${}^{5}D_{0}-{}^{7}F_{0}$ line is known to be
proportional to  $\alpha B_{20}^{2}$, where  $\alpha = 4/75\Delta_{20}$ 
and that of ${}^{5}D_{0}-{}^{7}F_{2}$ line is independent of $B_{20}$. 
Here, $\Delta_{20}$ is the energy separation between the  
${}^{7}F_{2}$  and ${}^{7}F_{0}$ states. Thus, the intensity ratio of 
${}^{5}D_{0}-{}^{7}F_{0}$ and ${}^{5}D_{0}-{}^{7}F_{2}$ transitions is 
given by 

\begin{equation}
\frac{I_{0-0}}{I_{0-2}}=\frac{4 B_{20}^{2}}{75\Delta_{20}^{2}}.
\end{equation}

\noindent
Taking $\Delta_{20}$ =900 cm$^{-1}$, variation of crystal field parameter, 
$B_{20}$, with increase in concentration of Eu$^{+3}$ is shown in Fig. 7.  
Its value  for 
Sample A 
and Sample B are estimated to be 1077 cm$^{-1}$ and 995 cm$^{-1}$, 
respectively. 
These values of $B_{20}$ are found to be in good agreement with the 
reported 
value 1260 cm$^{-1}$ for the commercial bulk system.\cite{35} Indirectly, 
it also 
reveals that the crystal field effect is similar in both Sample A and 
Sample B and match very well with the commercial bulk material. 


Fig. 8 shows both Stokes and anti-Stokes side of the Raman line at 375 
cm$^{-1}$, typical of cubic  Y$_{2}$O$_{3}$:Eu$^{3+}$ lattice. Local 
heating produced by the 
vibrational modes can give rise to a change in structure of the local 
environment of the Eu$^{3+}$ by causing ions to move from one 
equilibrium 
position to another. Stokes (I$_S$) and anti-Stokes(I$_{AS}$) Raman 
scattering intensities are related to a very good approximation 
by {\bf the relation\cite{42}}

\begin{equation}
n(\omega)I_{S} = [n(\omega)+1]I_{AS},
\end{equation}

\noindent
where $n(\omega)$ is the Bose-Einstein thermal factor, explicitly given 
by 

\noindent
\begin{equation}
n(\omega) = \frac{1}{exp({\hbar}\omega /k_{B}T)-1},
\end{equation}

\noindent
where $k_B$ is Boltzmann's constant and  $T$ {\bf is the equilibrium 
temperature of the sample.}
From 
the 
intensity ratio of the Stokes ($I_{s}$) and anti-Stokes ($I_{AS}$) Raman 
line one 
can calculate the equilibrium {\bf temperature} ($T$) of the material by 
using the equation\cite{42}

\begin{equation}
\frac{I_{s}}{I_{As}} = exp\left(\frac{\hbar\omega}{k_{B}T}\right),
\end{equation}

\noindent
where, $\omega$ is the phonon frequency. We have fitted both Stokes and 
anti-Stokes Raman spectra from the sample with Lorentzian line shape. In 
the non-linear least square fit the intensity, width, and the peak 
position have been kept as fitting parameters. From the intensity ratios 
of the Stokes and anti-Stokes Raman lines and using Eq. (10),  the 
equilibrium {\bf temperatures} of  Sample A and Sample B have been 
estimated  to 
be 247 K and 257 K, respectively. In other words, the 
material is stable till its 
equilibrium temperature is within this value. The equilibrium energy 
($k_{B}T$) of a 
system is a thermodynamic property.   Thus, we expect both Sample A and 
Sample B to have an equivalent local environment, which again is an 
important 
fact to be kept in mind while using  these nanocrystalline 
Y$_{2}$O$_{3}$:Eu$^{3+}$ in  nanotechnology.

\section{CONCLUSION}

In this article, we have reported optical properties of cubic 
nanocrystalline rare-earth doped inorganic oxide, 
Y$_{2}$O$_{3}$:Eu$^{3+}$, of average 
particle  size  15 nm. The electronic energy levels of this material are 
obtained from the photoluminescence measurements. The analysis of the 
variation in intensity of the luminescence spectrum at 612 nm for 
{\bf nanocrystalline} Y$_{2}$O$_{3}$:Eu$^{3+}$ as a 
function of Eu$^{3+}$ concentration in the host lattice leads to the 
observed 
fact that the dopant concentration of  4 at. wt $\%$ exhibits maximum 
intensity of the emission peak. We have discussed the 
crystalline nature of the 
material and the local field effect around the Europium ion, both of which 
influence the emission spectrum of this system. The equilibrium 
temperature, which affects the local field at the dopant 
site, has been estimated from the intensity ratio of Stokes and 
anti-Stokes 
Raman line. Table 1 summarizes the physical properties of the 
nanocrystalline Y$_{2}$O$_{3}$:Eu$^{3+}$, doped with 4 at. wt. $\%$ of 
Eu$^{+3}$ (Sample A), as obtained in our experiment and analysis. We claim 
that such nanocrystalline rare-earth doped oxides, 
with known 
characteristics can be used commercially as a red-emitting phosphor in 
nanotechnology.

\section*{Acknowledgement}
Authors thank P.V. Satyam and B. Satpati, 
IOP, Bhubaneshwar, India, for their assistance in the HRTEM work. Authors 
also thank Council of Scientific and Industrial Research, India, for 
financial support. AR thanks Department of Science and Technology, India 
for financial assistance.

\newpage
\noindent
$^{\mbox a)}$Author to whom correspondence should be addressed; 
Electronic mail: anushree@phy.iitkgp.ernet.in

\newpage

\noindent
{\bf Figure Captions}

\noindent
Fig. 1. (a)  HRTEM image of Sample A and (b) the frequency plot.

\vspace{0.2in}
\noindent
Fig. 2. X-ray diffraction pattern for Sample A (dotted line) and Sample B 
(solid line). 
\vspace{0.2in}

\noindent
Fig. 3. Emission spectrum of nanocrystalline Y$_{2}$O$_{3}$:Eu$^{3+}$
for $\lambda_{\mbox{exc}}$ = 488 nm. 
Intensities of (a) and (b) are not in the same scale.
\vspace{0.2in}

\noindent
Fig. 4. A schematic diagram of the energy level of Eu$^{3+}$ in 
nanocrystalline Y$_{2}$O$_{3}$:Eu$^{3+}$.

\vspace{0.2in}
\noindent
Fig. 5. Filled squares are the variation of the ratio of the intensities 
of  ${}^{5}D_{0}-{}^{7}F_{0}$ to  ${}^{5}D_{0}-{}^{7}F_{1}$ emission 
spectrum with at wt. $\%$ of Eu ion in 
nanocrystalline  Y$_{2}$O$_{3}$:Eu$^{3+}$. The solid line is the guide 
to the eye. The variation in intensity for ${}^{5}D_{0}-{}^{7}F_{2}$ 
emission peak with dopant concentration is shown in the inset.

\vspace{0.2in}
\noindent
Fig. 6. Emission spectrum of 612 nm line for bulk (solid  line) and 
nanocrystalline (dashed-dotted line) of  Y$_{2}$O$_{3}$:Eu$^{3+}$. The 
non-linear fit of the data for Sample A is shown in the inset.

\vspace{0.2in}
\noindent
Fig. 7. Variation of $B_{20}$ with concentration of Eu ion.

\vspace{0.2in}
\noindent
Fig. 8. Stokes and anti-Stokes Raman spectra for  Y$_{2}$O$_{3}$: 
Eu$^{3+}$. The squares are 
experimental data points. The solid lines are non-linear least square fit 
to the data.

\newpage
\begin{table}
\caption{Physical properties of nanocrystalline Y$_2$O$_3$:Eu$^{3+}$}
\begin{tabular}{|c|c|}\hline
{\bf Property} & {\bf Remark}\\\hline
Size of the particle & 15 nm \\
Size distribution & $\pm$ 5 nm\\
Phase & Cubic (corresponds to space group $Ia$3)\\
Intensity of the strongest emission line & Maximum for 4 at. wt. $\%$ 
doped Eu$^{3+}$ion\\
at 612 nm &\\
Crystal field parameter ($B_{20}$) & 1077 cm$^{-1}$\\
Equilibrium temperature & 247 K\\\hline
\end{tabular}
\end{table}


\vspace{0.25in}
\begin{thebibliography}{66} 
\bibitem{1} M. C. Brierley and P.W. France, Electron. Lett. {\bf 23}, 815 
(1987).
\bibitem{2}M. C. Brierley and P.W. France, Electron.   Lett. {\bf 24}, 539 
(1988).
\bibitem{3}J. L. Oomen, J. Lumin. {\bf 50}, 317 (1992).
\bibitem{4}G. Blasse and B.C. Grabmaier, Luminescent Materials (Springer, 
Berlin, 1994).
\bibitem{5}C. R. Ronda, J. Lumin. {\bf 72}, 49 (1997).
\bibitem{6}E. G. Behrens, F.M. Durville, R.C. Powell, and D. H. Blackburn, 
Phys. Rev. B {\bf 39}, 6076 (1989).
\bibitem{7}J. Silver, M. I. Martinez-Rubio, T.G. Ireland, G.R. Fern, and 
R.Withnall, J. Phys. Chem. B  {\bf 105}, 9107 (2001).
\bibitem{8}K.C. Mishra, J.K. Berkowitz, K.H. Johnson, and P.C. Schmidt, 
Phys. Rev. B  {\bf 45}, 10902 (1992).
\bibitem{9}E. Husson, C. Proust, P. Gillet, and J. P.Itié, Mater. Res. 
Bull. {\bf 34}, 2085 (1999).
\bibitem{10}M. Mitsunaga and N. Uesugi, J. Lumin. {\bf 48} $\&$ {\bf 49}, 
459 (1991).
\bibitem{11}R. G. Pappalardo and R. B. Hunt, Jr., J. Electrochem. Soc. 
{\bf 132}, 721 (1985).
\bibitem{12}D. B. M. Klassen, R. A.van Ham, and T. G. M. van Rijn, J. 
Lumin. {\bf 43}, 261 (1989).
\bibitem{13}A. Konard, J. Almansötter, J. Reichardt, A. Gahn, R. Tidecks, 
and K. Samwer, J. Appl. Phys. {\bf 3}, 1796 (1999).
\bibitem{14}W. van. Schaik and G. Blasse, Chem. Mater. {\bf 4}, 410 
(1992).
\bibitem{15}S. Erdei, R. Roy, G. Harshe, H. Juwhari, D. Agrawal, F. W. 
Ainger and W. B. White, Mater. Res. Bull. {\bf 30}, 745 (1995).
\bibitem{16}R. Schmechel, M. Kennedy, H. von  Seggern, H. Winkler, M. 
Kolbe, R. A. Fischer, Li Xaomao, A. Benker, M. Winterer, and H. Hahn, 
J. Appl. Phys. {\bf 89}, 1679 (2001).
\bibitem{17}A. Konard, T. Fries, A. Gahn , F. Kummer U. Herr, R. Tidecks, 
and K. Samwer,  J. Appl. Phys. {\bf 86}, 3129 (1999).
\bibitem{18}B.M.Walsh, N.P. Barnes, M. Petros, J. Yu, and U.N. Singh, J. 
Appl. Phys. {\bf 95}, 3255 (2004).
\bibitem{19}A. Pathak, S. Mahapatra, S. Mahapatra, S. Biswas, D. Dhak, 
N. Pramanik, A. Taraphder, and P. Pramanik, Am. Cer. Soc. Bull. 
{\bf 83}, 9301 (2004).
\bibitem{20}A. Roy and A.K. Sood, Phys. Rev. B {\bf 53}, 18 (1996).
\bibitem{21}X-ray Powder Diffraction, JCPDS file.
\bibitem{22}B. E. Warren, {\em X-ray Diffraction}, (Adison-Wesley, 
Reading, MA, 1969) p. 253 and 258; H. P. Klug and L.E. Alexander, 
{\em X-ray Diffraction Procedures 
for Polycrystalline and Amorphous Materials}, (Wiley, New York, 1973) p 
687 and 635; P. Scherrer, Go\"{tt}inger Nachichent {\bf 2}, 98 (1918).
\bibitem{23}A. K. Arora  and V.Umadevi, Appl. Spectrosc. {\bf 36}, 424 
(1982).
\bibitem{24}K. A. Gschneidner and L. Eyring, {\em Hand-book of Physics and 
Chemistry of Rare-Earths}, (Elsevier, Amsterdam/New York,1979).
\bibitem{25} N. C. Chang and J.B. Gruber, J. Chem. Phys. {\bf 41}, 3227 
(1964).
\bibitem{26}J. Silver, M.I. Martinez-Rubio, T.G. Ireland, and R.J. 
Withnall, J. Phys. Chem. B {\bf 106}, 7200 (2001).
\bibitem{27}M. G Paton and E. N Maslem, Acta. Crystallogr. Sec. B {\bf 
19}, 307 (1965); M. Faucher, Acta Crystallogr. Sec. B {\bf 36}, 3209 
(1980).
\bibitem{28}H. Ishibashi, K. Shimomoto, and K. Nakahighashi, J. Phys. 
Chem. Solids {\bf 9}, 809 (1994).
\bibitem{29}M. Mitric, B. Antic, M Balanda, D. Rodic and  M.L. Napijalo,  
J. Phys.; Condense Matter {\bf 9}, 4103 (1997).
\bibitem{30}B. R. Judd, Phys. Rev. {\bf 127}, 750 (1962).
\bibitem{31}G. S. Ofelt, J. Chem. Phys. {\bf 37}, 511 (1962).
\bibitem{32}S. Ram and S.K. Sinha, J. Solid State Chem. {\bf 66}, 225 
(1987).
\bibitem{33}A.F. Kirby and F.S. Richardson, J. Phys. Chem. {\bf 87}, 2557 
(1983).
\bibitem{34}W. C. Nieuwpoort and G. Blasse, Solid State Commun. {\bf 4}, 
227 (1966).
\bibitem{35}G. Nishimura and T. Kushida, Phys. Rev. B {\bf 37}, 9075 
(1988).
\bibitem{36}T. Tomiki et al., J. Phys. Soc. Jpn. {\bf 55}, 4543 (1986).
\bibitem{37}M. Buijs, A. Meyerink and G. Blasse, J. Lumin. {\bf 37}, 9 
(1987).
\bibitem{38}J. Heber,  K. H. Hellwege,  K. H. Kobler, U. Murmann, Z. Phys. 
{\bf 237}, 189 (1970).
\bibitem{39}H. Forest and G. Ban, J. Electrochem. Soc. {\bf 116},  474, 
(1969).
\bibitem{40}Y.-H. Tseng, B.-S. Chiou, C.-C. Peng, L. Ozawa, Thin Solid 
Films {\bf  330}, 173 (1998).
\bibitem{41}G. Nishimura and T. Kushida, Phys. Rev. B {\bf 52}, 4171 
(1995).
\bibitem{42}W. Hayes and R. Loudon, {\em Scattering of Light by Crystals}, 
(John Wiley $\&$ 
Sons, New York. Chichester, Brisbane, Toronto, 1978).
\end{thebibliography}
\end{document}